\documentclass[prl,twocolumn,floatfix,superscriptaddress]{revtex4-2}
\usepackage{graphicx}  
\usepackage{dcolumn}   
\usepackage{comment}
\usepackage{bm}        
\usepackage{amssymb}   
\usepackage{amsmath}
\usepackage{mathrsfs}
\usepackage{epigraph}
\usepackage{gensymb}
\usepackage{lmodern}
\usepackage{lipsum}
\usepackage{marvosym}
\usepackage{tipa}
\usepackage{bbold}
\usepackage{dsfont}
\usepackage{esint}
\usepackage{xcolor}
\usepackage{braket}
\usepackage{appendix}
\usepackage{natbib}
\usepackage{bbold}
\usepackage[colorlinks=true, linkcolor=red, urlcolor=blue,citecolor=blue]{hyperref}

\begin{document}

\title{Nonclassical dynamics of N\'{e}el vector and magnetization accompanied by  THz and high-harmonic radiation from ultrafast-light-driven NiO antiferromagnet insulator}

\author{Federico Garcia-Gaitan}
\affiliation{Department of Physics and Astronomy, University of Delaware, Newark DE 19716, USA}
\author{Adrian E. Feiguin}
\affiliation{Department of Physics, Northeastern University, Boston, MA 02115, USA}
\author{Branislav K.~Nikoli\'c}
\email{bnikolic@udel.edu}
\affiliation{Department of Physics and Astronomy, University of Delaware, Newark DE 19716, USA}

\begin{abstract}
Ultrafast-light-driven strongly correlated antiferromagnetic insulators, such as prototypical  NiO with large energy gap \mbox{$\simeq 4$ eV},  have recently attracted experimental attention using photons of both {\em above-gap} [K.  Gillmeister {\em et al.}, Nat. Commun. {\bf 11}, 4095 (2020)] and  {\em subgap}  [H. Qiu {\em et al.}, Nat. Phys. {\bf 17}, 388 (2021)] energy.  In the latter context, also of great interest to  applications, emission of THz radiation is also observed from NiO/Pt bilayers, where heavy metal (HM) Pt introduces strong spin-orbit coupling (SOC). However, in contrast to amply studied  spintronic THz emitters  using fs laser pulse (fsLP)-driven FM/HM (FM-ferromagnetic metal of conventional type, such as Fe, Ni, Co) bilayers, where ultrafast demagnetization takes place and is directly related to THz emission, microscopic mechanisms of electromagnetic (EM) radiation from NiO/HM bilayer remain obscure as total magnetization of NiO is {\em zero} prior to fsLP application. Here we employ the two-orbital  Hubbard-Hund-Heisenberg model and study, via numerically exact quantum many-body methods, the dynamics of its  N\'{e}el vector and {\em nonequilibrium} magnetization. This reveals {\em nonclassical} (i.e., not describable by Landau-Lifshitz  equation)  dynamics of N\'{e}el vector and nonequilibrium magnetization, changing {\em only} in length while not rotating, where the former is substantially reduced  in the case above-gap fsLP. Additionally, we compute  EM  radiation by time-dependence of  magnetization, as well as of local charge currents, finding that  both contributions   are significant in THz frequency range {\em only} in NiO with proximity SOC introduced by HM layer.  Above THz range, we find integer high-harmonic generation, as well as unusual {\em noninteger} harmonics for above-gap fsLP pump.
\end{abstract}
\maketitle

{\em Introduction.}---Pump-probe experiments~\cite{Wang2018e} with strongly correlated antiferromagnetic (AF) insulators (AFI), like prototypical NiO~\cite{Gillmeister2020}, reveal~\cite{Murakami2023} exotic  effects interweaving nonequilibrium many-body physics and quantum coherence that can persist on  surprisingly long timescales (such as \mbox{$\sim 1$ ps}~\cite{Gillmeister2020}) due to  a large Mott gap [Fig.~\ref{fig:fig2}] providing  protection against fast thermalization and heating.  The fs laser pulse (fsLP) in these experiments and typical theoretical studies~\cite{Gillmeister2020,Wang2017a} has a central frequency that is {\em above-gap} between two Hubbard bands [Fig.~\ref{fig:fig2}].  Theoretical interest also exists to understand quantum tunneling, multiphoton absorption and the so-called ``Keldysh crossover'' and the ensuing nonlinear doublon-holon pair production in the case of {\em subgap} fsLP pump~\cite{Oka2012,Shinjo2022,Murakami2018, Murakami2021}. Such  panoply of complex   nonequilibrium many-body states~\cite{Murakami2023}   cannot be found in fsLP-driven conventional band insulators and semiconductors, where single particle~\cite{Bajpai2019}  description suffices. 

\begin{figure}
    \centering
    \includegraphics[scale=0.11]{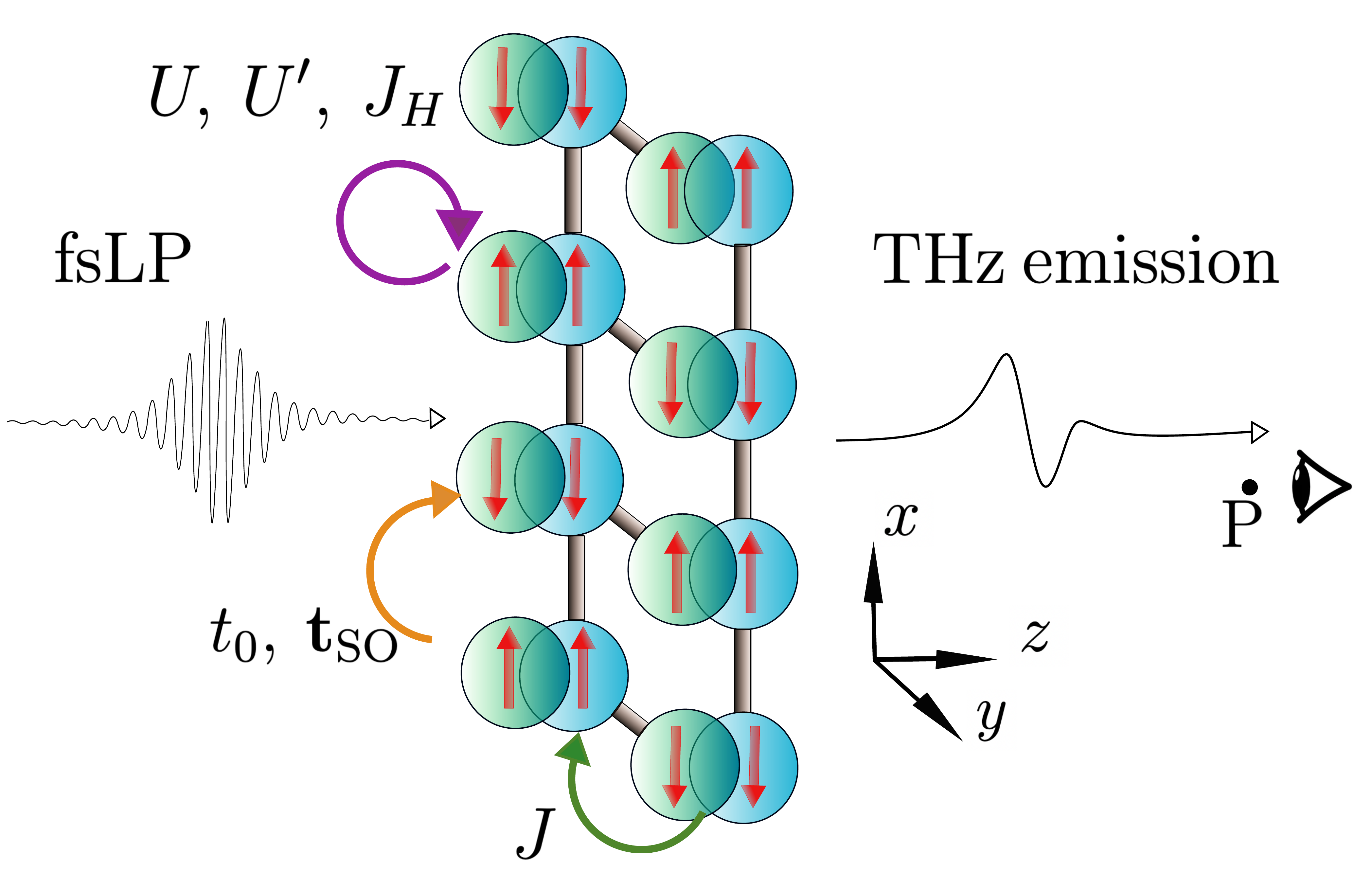}
    \caption{Schematic view of two-orbitals-per-site 2HHH model~\cite{Gillmeister2020} for NiO on a ladder geometry. It can also include additional Rashba SOC~\cite{Grytsyuk2016,Manchon2015}, introduced through $\mathbf{t}_\mathrm{SO}$ spin-dependent matrix~\cite{Nikolic2006} hoppings, and arising from proximity effects~\cite{Grytsyuk2016,MarmolejoTejada2017,Dolui2020b,Dolui2022,Zutic2019} in NiO/HM bilayers employed experimentally~\cite{Qiu2021,Rongione2023}. Other denoted parameters, describing local Coulomb and ($U$ and $U'$) Hund interactions ($J_H$), Heisenberg exchange interaction ($J$), and electron hopping ($t_0$) between sites, are explained in Eqs.~\eqref{eq:Hloc}---\eqref{eq:Htso}.  This setup is driven out of equilibrium by fsLP, and we compute EM radiation emitted by it in {\em both} THz and HHG frequency range.}
    \label{fig:fig1}
\end{figure}

\begin{figure}
    \centering
    \includegraphics[scale=0.72]{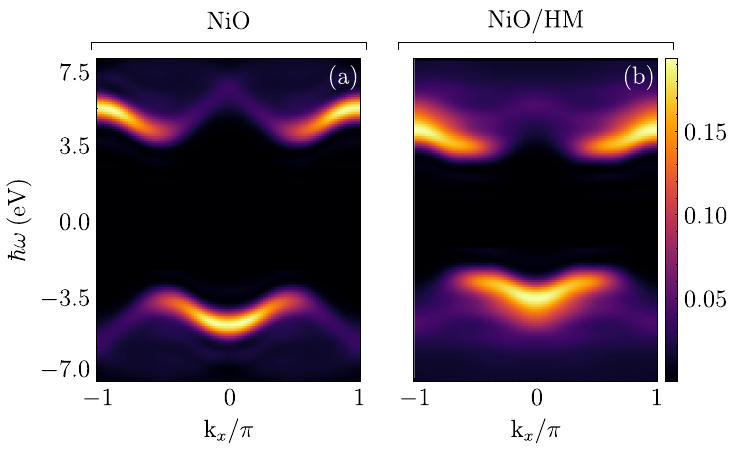}
\caption{The tDMRG-computed~\cite{Zawadzki2019,Yang2019} spectral function~\cite{Winkler2017a} of Hubbard model on a ladder [Fig.~\ref{fig:fig1}] for: (a) plain NiO; and (b) NiO with proximity induced Rashba SOC due to adjacent HM layer within NiO/HM bilayer  employed in THz spintronics experiments~\cite{Qiu2021,Rongione2023}. Here we focus on the bonding symmetry sector by choosing $k_y=0$~\cite{Yang2019}.}
    \label{fig:fig2}
\end{figure}

The same NiO material driven by subgap fsLP, but in combination with heavy metal (HM) layer (like Pt,W,Ta) introducing  effects~\cite{Grytsyuk2016} due to strong spin-orbit coupling (SOC), have been  very  recently explored~\cite{Qiu2021,Rongione2023} as  spintronic THz emitter~\cite{Seifert2016,Wu2017,Jungfleisch2018a,Rouzegar2022,Seifert2023,Leitenstorfer2023}. Isolated ferromagnetic metal (FM) layers (such as Co,Fe,Ni) or FM/HM bilayers have been intensely studied for nearly 30 years in order to understand ultrafast demagnetization~\cite{Beaurepaire1996,Chen2025}, as well as THz emission by these systems, which is much stronger in the case of bilayers~\cite{Seifert2016,Wu2017,Jungfleisch2018a,Rouzegar2022,Seifert2023,Liu2021} than in the case of a single FM layer~\cite{Beaurepaire2004,Rouzegar2022,Liu2021}. The fsLP in spintronic experiments typically has a central wavelength of $\simeq 800$ nm, so that its photons have energy centered around \mbox{$\hbar \Omega_0 \simeq 1.55$ eV} which is {\em subgap}~\cite{Qiu2021,Rongione2023,Wang2022} with respect to the separation \mbox{$\simeq 4$ eV} between the two Hubbard bands of NiO. The spintronic phenomena in such experiments on subgap-fsLP-excited NiO have been interpreted~\cite{Qiu2021,Rongione2023} by  borrowing the standard intuitive picture~\cite{Seifert2016,Wu2017,Jungfleisch2018a,Rouzegar2022,Seifert2023,Liu2021} (for its recent modifications, however, via microscopic theory see Refs.~\cite{Kefayati2024,Kefayati2024a,VarelaManjarres2024}) developed for FM/HM bilayers. That is, an ultrafast spin current is  somehow generated that flows from NiO into the HM layer, so that latter can   convert it into charge current via the inverse spin Hall effect~\cite{Saitoh2006}. The time-dependent charge current is needed to obtain sizable electromagnetic (EM) radiation in the far-field (FF) region [Fig.~\ref{fig:fig1}], as well as to interpret~\cite{Qiu2021,Rongione2023} the enhancement~\cite{Rouzegar2022} of the emitted radiation when switching from FM to FM/HM systems. This is because magnetic dipole radiation~\cite{Beaurepaire2004,Rouzegar2022,Liu2021} from  time-dependent magnetization $\mathbf{M}(t)$  is $1/c$ smaller~\cite{Kefayati2024}  than radiation by a time-dependent charge current.   However, this picture does not explain the microscopic  mechanism that  generates the assumed spin current in NiO/HM bilayers (only speculation exist thus far~\cite{Han2023}), or why the frequency spectrum of such current has features within \mbox{0.1--3 THz} range that is imprinted in the EM radiation  detected experimentally~\cite{Qiu2021,Rongione2023}. It is obvious that fsLP will drive electrons into dynamics at its own center frequency $\Omega_0$, as well as  at {\em integer}  (typically odd~\cite{Lange2024,TancogneDejean2022}) multiples of  $\Omega_0$  for sufficiently intense fsLP. Such high-harmonic generation (HHG) in current and EM radiation has been vigorously explored in recent years in diverse  quantum materials~\cite{Ghimire2018,Schmid2021},  including strongly correlated ones~\cite{Murakami2018,Murakami2021,Imai2020,Orthodoxou2021,Lange2024,Murakami2023}. Finally, it remains unclear what type of dynamics is obeyed by the N\'{e}el vector (as the difference of two sublattices  magnetizations) and magnetization [as the sum of sublattice magnetizations, which is  necessarily a nonequilibrium quantity because $\mathbf{M}(t=0) \equiv 0$], when compared to standard demagnetization~\cite{Beaurepaire1996,Chen2025,Rouzegar2022} of FM layers, where magnetization vector shrinks.  In the case of thin FM layers, a rapid and straightforward analysis of  the  direction of $\mathbf{M}(t)$ and its magnitude  is achieved via~\cite{Rouzegar2022}  magneto-optical Kerr or Faraday effects. In contrast, they do not apply to AFs, so novel ideas have been explored~\cite{Saidl2017} to detect the presumed~\cite{Galkina2021,Gomonay2010,Rongione2023} rotation of the N\'{e}el vector. 

Thus, developing a microscopic understanding of the response of AFI   to subgap fsLP (that is, by starting from a suitable quantum many-body Hamiltonian and using tools of nonequilibrium quantum statistical mechanics),  would also help to resolve a number of outstanding issues  in AF optospintronics~\cite{Nemec2018}. Note that, specifically for NiO which is a strongly correlated material  sharing features of both Mott and charge-transfer insulators~\cite{Gillmeister2020,Lechermann2019}, X-ray techniques applied as a probe after subgap fsLP pump have revealed~\cite{Wang2022} possible substantial~\cite{Wang2018e,Wang2017a} changes of its electronic structure, such as  emergence  of midgap states and Hubbard gap reduction persisting on time scales \mbox{$> 2$ ps}. Such phenomena originating from  charge dynamics must be taken into account together with local spin dynamics, as they can lead to  {\em inextricable}  complex spin-charge  dynamics~\cite{Gillmeister2020,Bittner2018}. In the case of weakly correlated FMs, proper description of spin-charge dynamics is achieved via time-dependent density functional theory (TDDFT) which has  provided~\cite{Krieger2015,Chen2019a,Wu2024,Mrudul2024,Shokeen2017} a most detailed  insight into a sequence of fast changing events~\cite{Tengdin2018} and their effect~\cite{Kefayati2024,Kefayati2024a} on THz emission. However,  application~\cite{Dewhurst2018} of TDDFT to NiO is impeded by intricacies in  including a time-dependent~\cite{TancogneDejean2018,Orhan2019,Murakami2023} Hubbard $U$  to properly capture a strong  on-site Coulomb interaction that is also dynamical (as opposed to static $U$ in conventional equilibrium DFT+U~\cite{Anisimov1991} calculations).  

In this Letter, we aim to capture all the essential physics of strongly correlated electrons within NiO, in the presence of fsLP and SOC,  by employing a two-orbital  Hubbard-Hund-Heisenberg (2HHH)  model~\cite{Gillmeister2020} and by simulating its time evolution via numerically exact quantum-many body  methods. Its ground AF state  is formed by local spins $S=1$~\cite{Nag2020} at each site,  that are  composed of two elemental spins $s = 1/2$ located on two  orbitals at that site [Fig.~\ref{fig:fig1}], with Hund interaction included between them in order to achieve fixed and stable $S$ per atom~\cite{Hoffmann2020,Stadler2021}.  In addition, to take into account adjacent nonmagnetic HM  layer in THz spintronics experiments~\cite{Qiu2021,Rongione2023}, we also include SOC of the Rashba type~\cite{Grytsyuk2016,Manchon2015} into the 2HHH model of Ref.~\cite{Gillmeister2020}. Thus, the introduced SOC models proximity effects~\cite{Grytsyuk2016,MarmolejoTejada2017,Dolui2020b,Dolui2022,Zutic2019} around the NiO/HM interface, that modifies the electronic structure on the NiO side due to the HM layer (NiO, in turn, modifies bands of HM layer, but we do not include HM layer explicitly due to high computational expense). This model is placed onto a ladder geometry  [Fig.~\ref{fig:fig1}], allowing for its spectral function~\cite{Zawadzki2019,Yang2019,Winkler2017a} 
to be computed via numerically (quasi)exact simulations using time-dependent density matrix renormalization group (tDMRG)~\cite{White2004,Daley2004,Schmitteckert2004,Feiguin2011} applicable to quasi-one-dimensional lattices. We complement the study with an additional set of numerical simulations using massively parallel exact diagonalization (ED) methods~\cite{Innerberger2020} for Hubbard-type systems, implemented within the H$\Phi$ package~\cite{Ido2024,Kawamura2017}. 
This technique allows us to access longer times (required for THz radiation calculations) than those possible via tensor network algorithms (like tDMRG) encountering ``entanglement barrier''~\cite{Lerose2023,Garciagaitan2024}. 

\begin{figure*}[th!]
    \centering
    \includegraphics[width=\textwidth]
    {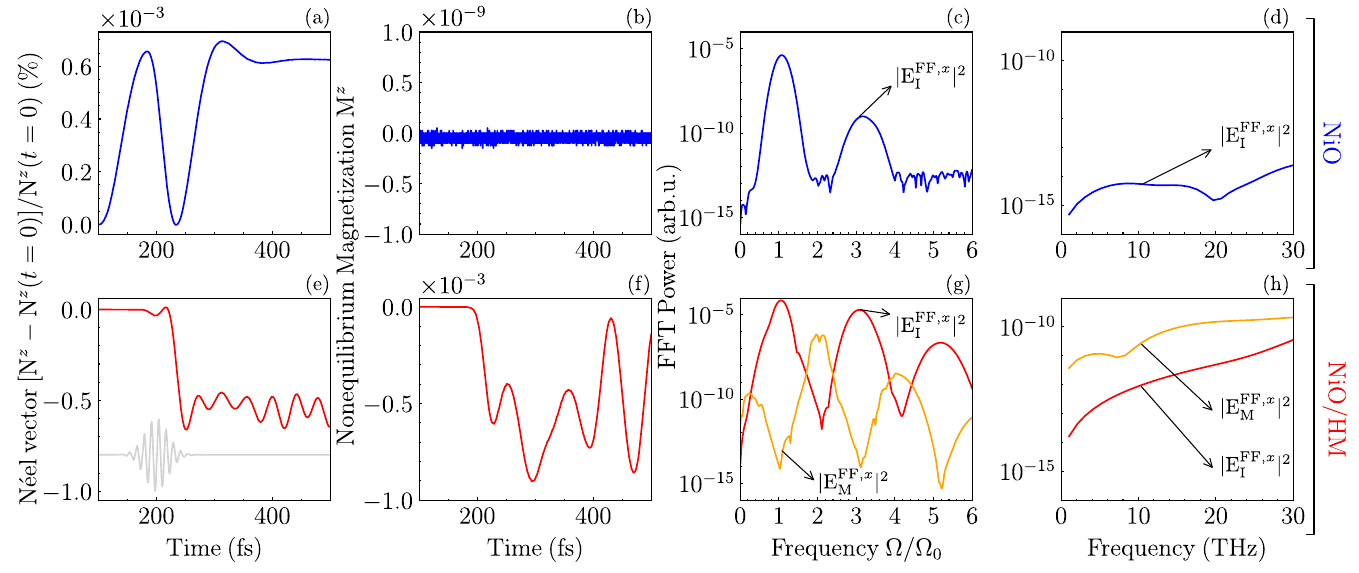}
    \caption{Time dependence,  initiated by a fsLP with {\em subgap} central frequency, of: (a),(e)  N\'eel vector; and (b),(f) nonequilibrium magnetization. (c),(d),(g),(h) Fast Fourier transform (FFT) power spectrum---within (c),(g) HHG frequencies or or (d),(h) THz frequencies---of the $x$-component of electric field of EM  radiation emitted by the dynamics of magnetization [Eq.~\eqref{eq:chargefield}] 
    or bond charge currents  [Eq.~\eqref{eq:chargefield}]. The top row of panels is for plain NiO, while the bottom row is for NiO assumed to contain~\cite{Grytsyuk2016}  the Rashba SOC due to proximity efects~\cite{MarmolejoTejada2017,Zutic2019,Dolui2020b,Dolui2022} from  HM layer within NiO/HM bilayer studied in recent  experiments~\cite{Qiu2021,Rongione2023}.} 
    \label{fig:fig3}
\end{figure*}

\begin{figure*}
    \centering
    \includegraphics[width=\textwidth]
    {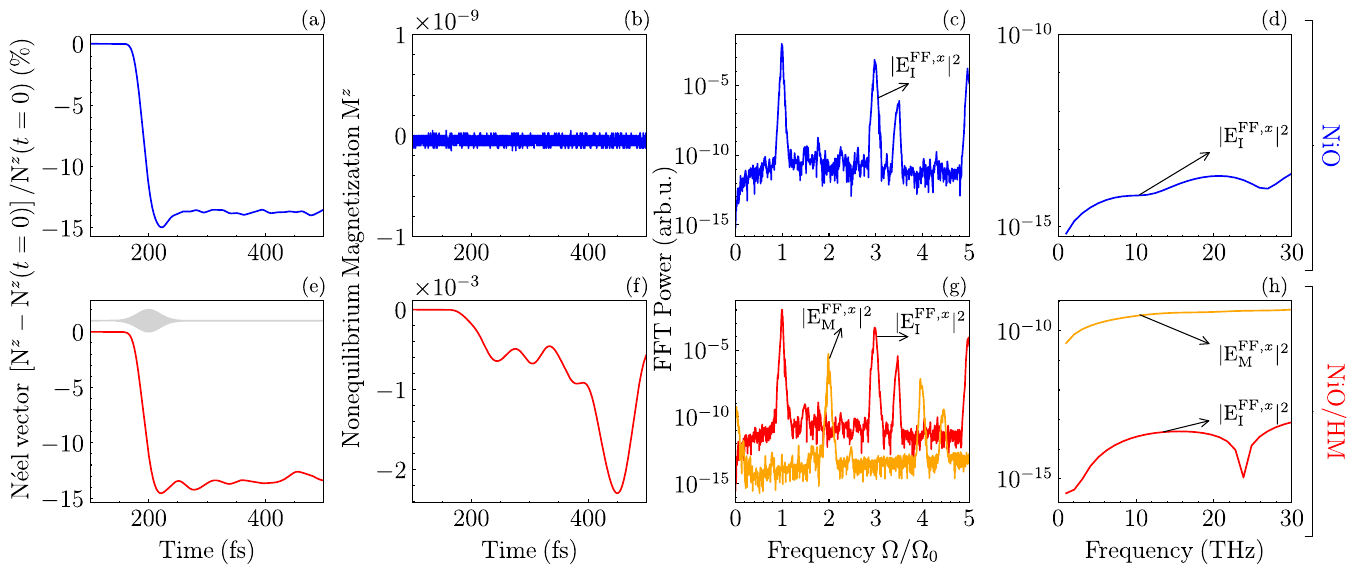}
    \caption{The same information as in Fig.~\ref{fig:fig3}, but using fsLP whose central frequency of \mbox{$\hbar \Omega_0 = 8$ eV} is {\em above-gap} of  Fig.~\ref{fig:fig2}.}
    \label{fig:fig4}
\end{figure*}

By considering a multi-orbital ladder, spin-charge separation is largely suppressed, so  a plethora of dissipative mechanisms based on electron-spin interaction become active, such as the AF background interaction with doublons and holons~\cite{Gillmeister2020, Murakami2023} or local dissipation caused by the Hund interaction~\cite{Rincon2018, Lysne2020b, Strand2017}. Our principal results, revealing highly nonclassical dynamics of N\'{e}el vector and nonequilibrium magnetization, and the ensuing radiation at both HHG of fsLP and in much lower THz range, are given in Figs.~\ref{fig:fig3} and ~\ref{fig:fig4}. Prior to discussing these results, we introduce useful concepts and notations.

{\em Model and Methods.}---The monolayer~\footnote{We use monolayer of NiO and  do not include HM
layer explicitly due to high computational expense of our numerically exact calculations. Nevertheless, this limitation is not a drawback for capturing  essential features of experiments as they find~\cite{Rongione2023} the highest intensity of emitted THz radiation for the  thinnest  NiO layer driven by subgap fsLP.}  of  NiO is modeled on a tight-binding ladder [Fig.~\ref{fig:fig1}] with two orbitals per site hosting Ni only, while O atom is not modeled explicitly albeit O-mediated interactions areincluded.  The same 2HHH model, but without any SOC, was used in Ref.~\cite{Gillmeister2020} where above-gap pumping of NiO by fsLP was studied experimentally and theoretically. Our Hamiltonian, \mbox{$\hat{H} = \hat{H}_{\rm local} + \hat{H}_{\rm ex} + \hat{H}_{\rm TB} + \hat{H}_{\rm SOC}$}, is built on top of the model from Ref.~\cite{Gillmeister2020} by  including  possible Rashba SOC~\cite{Grytsyuk2016,Manchon2015} to describe the NiO/HM  bilayer used in THz spintronic  experiments~\cite{Qiu2021,Rongione2023}. The local terms of $\hat{H}_\mathrm{local}$ account for the Hubbard and Hund physics
\begin{eqnarray}\label{eq:Hloc}
    \hat{H}_{\rm local} & =& U\sum_{i,\alpha}\hat{n}_{i,\alpha \uparrow}\hat{n}_{i,\alpha \downarrow} - \mu \sum_{i,\alpha, \sigma}\hat{n}_{i\alpha\sigma} -g\mu_BB_z^{\rm imp}\hat{s}_{1\alpha}^z  \nonumber \\ 
    && + \sum_{i,\alpha < \beta}\sum_{\sigma, \sigma'}(U'-J_{\rm H}\delta_{\sigma\sigma'})\hat{n}_{i\alpha\sigma}\hat{n}_{i\beta\sigma'} \nonumber \\ 
    && + \gamma J_{\rm H}\sum_{i,\alpha \neq\beta}\left( \hat{c}_{i\alpha\uparrow}^\dagger \hat{c}_{i\alpha\downarrow}^\dagger \hat{c}_{i\beta\downarrow} \hat{c}_{i\beta\uparrow} + \mathrm{H.c.} \right) \nonumber \\
    && +\gamma J_{\rm H} \sum_{i, \alpha\neq \beta}\left(\hat{c}_{i\alpha\uparrow}^\dagger \hat{c}_{i\beta\downarrow}^\dagger \hat{c}_{i\alpha\downarrow} \hat{c}_{i\beta\uparrow} +\mathrm{H.c.}\right).
\end{eqnarray}
Here $\hat{c}_{i\alpha \sigma}$ ($\hat{c}_{i\alpha \sigma}^\dagger$) is the creation (annihilation) operator of an electron of spin $\sigma=\uparrow,\downarrow$ in  orbital $\alpha=1,2$ located at site~$i$;  $\hat{n}_{i\alpha\sigma}$ is the corresponding number operator; $U$, $U'$, and $J_{\rm H}$ are the intra-orbital Coulomb, inter-orbital Coulomb, and inter-orbital (or Hund~\cite{Hoffmann2020,Stadler2021}) exchange interaction, respectively; $\mu$ is the onsite chemical potential;  and $B_z^\mathrm{imp}$ is a magnetic field added~\cite{Petrovic2021b} on site $i=1$ to lift the degeneracy between spin-$\uparrow$ and spin-$\downarrow$ electrons. 
The Heisenberg exchange interaction between spins at nearest neighbor (NN) sites is given by 
\begin{equation}\label{eq:Hex}
\hat{H}_\mathrm{ex} = \sum_{\braket{ij},\alpha} \bigg[J\left (  \hat{s}^x_{i\alpha}\cdot \hat{s}^x_{j\alpha} + \hat{s}^y_{i\alpha}\cdot \hat{s}^y_{j\alpha} \right ) + J_{z}\hat{s}^z_{i\alpha}\cdot \hat{s}^z_{j\alpha}\bigg], 
\end{equation}
where $\braket{ij}$ denotes summation over the NN sites;   \mbox{$\hat{s}_{i\alpha}^p= \sum_{\sigma,\sigma'} \hat{c}_{i\alpha\sigma}^\dagger \frac{1}{2}\hat{\sigma}_{\sigma\sigma'}^p\hat{c}_{i\alpha\sigma'}$} is the electron spin operator expressed using  $\hat{\sigma}^p$ as one of three ($p=x,y,z$) Pauli matrices; and \mbox{$J=J_z=0.1~\rm eV$} in the  isotropic case. The kinetic term in tight-binding (TB) Hamiltonian  
\begin{equation} \label{eq:Hkin}
\hat{H}_\mathrm{TB} = -t_0 \sum_{\braket{ij},\alpha,\sigma} \left( \hat{c}_{i\alpha\sigma}^\dagger \hat{c}_{j\alpha\sigma} + \mathrm{H.c.} \right),
\end{equation}
where $t_0$ is the hopping parameter. An additional TB term is employed to introduce proximity SOC---switched off in NiO case and switched on in NiO/HM case in Figs.~\ref{fig:fig3} and ~\ref{fig:fig4}---as given by 
\begin{equation}\label{eq:Htso}
\hat{H}_\mathrm{\rm SOC} = \sum_{\braket{ij},\alpha}\left(\hat{\mathbf{c}}^\dagger_{i\alpha} \mathbf{t}_{\mathrm{SO}} \hat{\mathbf{c}}_{j\alpha} + \mathrm{H.c.} \right),
\end{equation}
where $\hat{\mathbf{c}}_{i\alpha}^\dagger = (\hat{c}_{i\alpha\uparrow}^\dagger \ \  \hat{c}_{i\alpha\downarrow}^\dagger)$ denotes row vector of two creation operators and $\mathbf{t}_{\mathrm{SO}}$ is a direction-dependent $2\times 2$ matrix hopping~\cite{Nikolic2006} with values \mbox{$-it_{\mathrm{SO}}\hat{\sigma}_y (it_{\mathrm{SO}}\hat{\sigma}_x)$} for horizontal (vertical) bonds. 
The $\hat{H}_\mathrm{\rm SOC}$ term 
is of the Rashba type~\cite{Manchon2015}, assumed to originate from proximity to HM layer employed experimentally~\cite{Qiu2021,Rongione2023}, as found in DFT calculations on FM/HM~\cite{Grytsyuk2016} of AFI/HM~\cite{Dolui2022} bilayers. Realistic parameters values for NiO are taken from prior first-principles calculations for strongly correlated electrons~\cite{Held2007}, such as from DFT+U~\cite{Anisimov1991} and/or  DFT+dynamical mean field theory~\cite{Kunes2007,Zhang2019} studies: \mbox{$U\approx 8~\rm eV$}; \mbox{$t_0\approx 1$ eV};  \mbox{$J_{\rm H}\approx1~\rm eV$}; and we set \mbox{$U' = U-2J_{\rm H}$} and \mbox{$\gamma=1$} for symmetry reasons~\cite{Rincon2018}. Half-filling is selected by setting the chemical potential to \mbox{$\mu=(3U-5J_H)/2$}~\cite{Gillmeister2020}. The magnetic field at site $1$, \mbox{$g\mu_BB_z^\mathrm{imp}=0.1~\rm eV$}, as generated by, e.g., a static impurity, induces~\cite{Petrovic2021b}  N\'{e}el ``checkerboard'' order $\langle \hat{S}_{i}^z \rangle = - \langle \hat{S}_{j}^z \rangle \neq 0$ ($i$ and $j$ are two NN sites) in the ground state (GS). Nevertheless, the GS in our simulations  retains nonzero entanglement, as witnessed in many recent experiments even at finite temperature ~\cite{Scheie2021,Laurell2024}, so it is not identical to unentangled N\'{e}el ket $\ket{\uparrow \downarrow \ldots \uparrow \downarrow}$.  

The fsLP is introduced via its vector potential [gray line in Figs.~\ref{fig:fig3}(e) and ~\ref{fig:fig4}(e)] of amplitude $A_\mathrm{max}$, which couples to electrons in the form of a Peierls phase~\cite{Panati2003,Li2020} multiplying hoppings in Eqs.~\eqref{eq:Hkin} and ~\eqref{eq:Htso} by a factor \mbox{$P=\exp \bigg [\textit{i}z_\mathrm{max} e^{\frac{-(t-t_p)^2}{2\sigma_\mathrm{light}^2}}\cos(\Omega_0 t) \bigg ]$}, so  $t_0(t)=Pt_0$ and \mbox{$\mathbf{t}_{\mathrm{SO}}(t)=P\mathbf{t}_{\mathrm{SO}}$}. Here  \mbox{$z_\mathrm{max} = e a_0 A_\mathrm{max}/\hbar = 0.2$} is the dimensionless parameter~\cite{Bajpai2019} quantifying the fsLP intensity;  $a_0$ is the lattice constant; and \mbox{$\sigma_\mathrm{light} = 20$ fs} determines the width of the Gaussian envelope which is initially centered at \mbox{$t_p=200$ fs}. The center frequency of the fsLP is either \mbox{$\hbar \Omega_0=1.55$ eV} in Fig.~\ref{fig:fig3}, corresponding to a subgap \mbox{$800$ nm} wavelength commonly employed in THz spintronic experiments~\cite{Qiu2021,Seifert2016,Wu2017,Jungfleisch2018a,Rouzegar2022,Seifert2023,Liu2021,Rongione2023}; or \mbox{$\hbar \Omega_0=8$ eV} in Fig.~\ref{fig:fig4} for above-gap fsLP [gaps are shown in Fig.~\ref{fig:fig2}].  The electric field of the fsLP is linearly polarized along the $xy$-direction (i.e., diagonal to the ladder) in Fig.~\ref{fig:fig1}. The time evolution described by $\hat{H}(t)$ on a $4\times 2$ ladder is obtained via ED~\cite{Innerberger2020} by means of the H$\Phi$ package~\cite{Kawamura2017,Ido2024}, where the GS was found using the Lanczos algorithm and the evolution operator is computed via a Taylor expansion considering up to $15$ terms. The time step was chosen as $\delta t =0.005 \hbar/t_0$.

To compute the EM radiation generated by the charge dynamics, the expectation value $I_{ij} \equiv \langle \hat{I}_{ij} \rangle$ of the bond charge current operator~\cite{Nikolic2006}, 
   \mbox{$\hat{I}_{ij} = \frac{ie}{\hbar} \sum_{\alpha} \big[ \hat{\mathbf{c}}_{i\alpha}^\dagger \{t_0(t) \hat{\sigma}_0 + \mathbf{t}_{\mathrm{SO}}(t)\}\hat{\mathbf{c}}_{j\alpha}-\mathrm{H.c.} \big]$} 
from site $i$ to site $j$ with  $\hat{\sigma}_0$ being a unit $2\times2$ matrix,  is plugged~\cite{Ridley2021,Suresh2023,VarelaManjarres2024} into the Jefimenko formula~\cite{Jefimenko1966} for the electric field
\begin{eqnarray}\label{eq:chargefield}
\mathbf{E}^\mathrm{FF}_{I}(\mathbf{r}, t) & = & \frac{1}{4\pi \epsilon_0 c^2}\sum_{P_{i \rightarrow j}=1}^{N_b} \int_{P_{i \rightarrow j}}\bigg[ (\mathbf{r}-\mathbf{l})\frac{\partial_tI_{ij}(t_r)}{|\mathbf{r}-\mathbf{l}|^3}(\mathbf{r}-\mathbf{l})\cdot \mathbf{e}_x  \nonumber \\
&& 
\mbox{} - \frac{\partial_t I_{ij}\left(t_r\right)}{|\mathbf{r}-\mathbf{l}|} \mathbf{e}_x \bigg ] d l.
\end{eqnarray}
The most general Jefimenko formula~\cite{Jefimenko1966}, as  proper solution of the Maxwell equations~\cite{Griffiths1991}, is  reorganized~\cite{McDonald1997} above in order to isolate the FF contributions decaying as $\sim 1/r$. We compute radiation  at point P in Fig.~\ref{fig:fig1} which is at a distance $100a_0$ away from NiO. Here,  $t_r=t-|\mathbf{r}-\mathbf{l}|/c$ emphasizes  retardation due to relativistic causality;  bond currents are assumed to be homogeneous~\cite{Ridley2021,Suresh2023,VarelaManjarres2024} along the path $P_{i\rightarrow j}$ from site $i$ to site $j$ of length $dl$; $N_b$ is the number of bonds; and we use shorthand notation $\partial_t \equiv \partial/\partial_t$. In addition, we also compute the electric field of FF radiation by magnetic dipole, i.e.,  due to the time-dependence of the nonequilibrium   magnetization 
\begin{equation}
\label{eq:dipolefield}
    \mathbf{E}^{\mathrm{FF}}_{M} (\mathbf{r}, t) = \frac{1}{4\pi\epsilon_0 c^3}\sum_i \frac{\mathbf{r}-\mathbf{l}_i}{|\mathbf{r}-\mathbf{l}_i|^2}\times \partial^2_t\mathbf{M}_i(t_r),
\end{equation}
where $\mathbf{l}_i$ indicates the location of site $i$. Time-dependent magnetization $M^z(t)$ and N\'{e}el vector $N^z(t)$ are obtained by summing spin expectation values at each site, \mbox{$\mathrm{M}^z=\sum_i M_i^z=\sum_{i,\alpha}\langle \hat{s}_{i\alpha}^z\rangle(t)$} and \mbox{$\mathrm{N}^z=\sum_i N_i^z \sum_{i,\alpha} (-1)^{i}\langle \hat{s}_{i\alpha}^z\rangle(t)$}, where  a ``snake-like'' enumeration of  ladder sites is assumed. Note that  other (the $x$- and $y$-components) are vanishingly small. 


%

{\em Results and Discussion.}---The usual first take at     interpreting  experiments on subgap light-driven AFIs, including NiO~\cite{Kampfrath2011,Satoh2010,Rongione2023},  invokes~\cite{Kampfrath2011,Kimel2009,Blank2023,Satoh2010}   
a direct coupling of light magnetic field (or an effective one due to inverse Faraday~\cite{Satoh2010} or inverse Cotton-Mouton~\cite{Rongione2023} effects) to local magnetization of AFIs. This leads to classical dynamics of the N\'{e}el vector, which rotates without changing its length~\cite{Rongione2023}, in  accord with  phenomenological theories~\cite{Galkina2021,Gomonay2010} of Landau-Lifshitz (LL) type ~\cite{Landau1935}. However, limitations of this approach 
are often found in experiments~\cite{Satoh2010,Formisano2024},  which is not surprising as light-charge coupling is much stronger~\cite{Chen2019a}, so electrons should be explicitly included. But the picture of classical LL dynamics~\cite{Galkina2021,Gomonay2010,Rongione2023} is appealing because it is difficult to develop intuition on how lectrons of AFI, with gapped energy spectrum [Fig.~\ref{fig:fig2}],  absorb subgap light and then  affect its magnetic ordering. Ie case of AF metals, it is  easy to envisage (and calculate~\cite{Suresh2023}) howthat  fsLP generates photocurrent of conduction electrons, which are then spin-polarized by the magnetic background and  exert spin torque~\cite{Suresh2023} onto local magnetization. Its dynamics follows  (for weak-intensity  fsLP to avoid demagnetization~\cite{Beaurepaire1996,Chen2025})  classical LL dynamics 
(valid at sufficiently high temperature and for AFIs with $S>1$~\cite{Garciagaitan2024}). For AFIs, there is no such shortcut and  one has to handle complexities of  photoexcited Hubbard model~\cite{Oka2012,Shinjo2022,Murakami2018, Murakami2021,Wang2017a},  which cannot ~\cite{Mentink2015},  in general,   be reduced to studying just their final outcome on magnetic order. 

Indeed, our quantum many-body calculations for subgap fsLP-driven NiO reveal highly {\em nonclassical}~\cite{Petrovic2021b,Mitrofanov2021} (i.e., outside any   description by  LL-type theories~\cite{Galkina2021,Rongione2023}) dynamics of the N\'eel vector and magnetization in Fig.~\ref{fig:fig3}. That is, both vectors are changing length along the $z$-axis (orthogonal to the ladder in Fig.~\ref{fig:fig1}) while {\em not rotating} at all. Nonequilibrium magnetization remains zero in plain NiO [Fig.~\ref{fig:fig3}(b)]. However,  $M^z(t) \neq 0$ in SO-proximitized NiO  [Fig.~\ref{fig:fig3}(f)], which is akin to experimentally observed~\cite{Afanasiev2019} weak ferromagnetism in photodoped Mott insulator  with native SOC. The  nonzero $\partial_t^2 M^z(t_r)$ emits magnetic dipole [Eq.~\eqref{eq:dipolefield}] THz radiation [orange line in Fig.~\ref{fig:fig3}(h)], which can (surprisingly,  when compared to FM layers~\cite{Kefayati2024}) surpass the contribution from $\partial_t I_{ij}(t_r)$ [red curve in Fig.~\ref{fig:fig3}(h)].  Importantly, THz emission from both of these  two sources is significant {\em only} only when proximity SOC is present in NiO/HM bilayer [Fig.~\ref{fig:fig3}(h)], in full accord with  experiments~\cite{Qiu2021,Rongione2023}. Thus, our theory explains these experiments  without invoking qualitative speculations~\cite{Qiu2021,Han2023,Rongione2023}, while showing that concepts borrowed from FM/HM systems (like interlayer spin current and spin-to-charge conversion within HM~\cite{Seifert2016,Wu2017,Rouzegar2022,Seifert2023,Liu2021,Kefayati2024a}) are not necessary for THz emission from AFI. The  magnetic dipole radiation exhibits even integer HHG [Fig.~\ref{fig:fig3}(g)], while odd ones are expected~\cite{Lange2024} for radiation from $\partial_t I_{ij}(t_r)$, as dictated by symmetry-imposed selection rules of Floquet group theory~\cite{Neufeld2019}. Although SOC breaks inversion symmetry, odd integer HHG is preserved~\cite{Lysne2020,TancogneDejean2022}. This offers a scheme---{\em detect  even HHG in EM radiation from AFI}---which directly corroborates how magnetization dynamics in  magnets driven by fsLP can be much faster~\cite{Kimel2019,Mrudul2024} than observed in low-energy transport experiments~\cite{Han2023}.

For above-gap fsLP in Fig.~\ref{fig:fig4}, we find reduction of  N\'{e}el vector by up to 15\%, signifying suppression~\cite{Wang2017a} of AF order in the GS. This is in sharp contrast to its minuscule change for subgap fsLP in Fig.~\ref{fig:fig3}. Similarly to subgap fsLP, nonequilibrium magnetization [Fig.~\ref{fig:fig4}(f)] emerges {\em only} when proximity SOC in switched on.  Unlike the subgap fsLP case, charge current and its THz radiation are not significantly enhanced by including proximity SOC [Fig.~\ref{fig:fig4}(d) vs. Fig.~\ref{fig:fig4}(h)]. The selection rules for HHG remain the same, allowing only odd integer HHG of $E^{\mathrm{FF},x}_I$ [Fig.~\ref{fig:fig4}(c),(g)] and even ones of $E^{\mathrm{FF},x}_M$ [Fig.~\ref{fig:fig4}(g)]. Curiously, we also find {\em noninteger} HHG of both $E^{\mathrm{FF},x}_I$ and $E^{\mathrm{FF},x}_M$, which are beyond standard Floquet theory and its selection rules~\cite{Neufeld2019}. Theoretical~\cite{Lange2024} and experimental~\cite{Schmid2021} reports of noninteger HHG are scarce, and their understanding is in its infancy. For example, they could arise~\cite{Lange2024} from correlation-driven population of multiple Floquet states (unlike population of a single state used in Floquet group theory~\cite{Neufeld2019}), whose exploration we relegate to future studies.

\begin{acknowledgments}
		F.G-G. and B.K.N. were supported by the US National Science Foundation (NSF)  through the  University of Delaware Materials Research Science and Engineering Center, DMR-2011824. The supercomputing time was provided by DARWIN (Delaware Advanced Research Workforce and Innovation Network), which is supported by NSF Grant No. MRI-1919839. A.E.F. was suppported by the  US Department of Energy, Office of Basic Energy Sciences under Grant No. DE-SC0022311.
	\end{acknowledgments}

\bibliography{references}


 \end{document}